\documentclass{article}

\usepackage{graphicx}

\author{Aron C. Wall\footnote{aronwall@umd.edu}
\\ \textit{Maryland Center for Fundamental Physics} \\ \textit{Department of Physics} \\ \textit{University of Maryland} \\ \textit{College Park, MD 20740-4111, USA} }
\title{Testing the Generalized Second Law \\ in 1+1 dimensional Conformal Vacua: \\ An Argument for the Causal Horizon.}

\date{\today}

\begin{document}

\maketitle

\begin{abstract}
The anomalous conformal transformation law of the generalized entropy is found for dilaton gravity coupled to a 1+1 conformal matter sector with central charges $c = \tilde{c}$.  (When $c \ne \tilde{c}$ the generalized entropy is not invariant under local Lorentz boosts.)  It is shown that a certain second null derivative of the entropy, $S^{\prime\prime}_\mathrm{gen} + (6/c)(S^\prime_\mathrm{out})^2$, is primary, and therefore retains its sign under a general conformal transformation.  Consequently all conformal vacua have increasing entropy on causal horizons.  Alternative definitions of the horizon, including apparent or dynamical horizons, can have decreasing entropy in any dimension $D \ge 2$.  This indicates that the generalized second law should be defined using the causal horizon.
\end{abstract}

\section{Introduction}\label{intro}

This article will test the generalized second law (GSL) for 1+1 dilaton gravity, semiclassically coupled to a conformal field theory (CFT).  The generalized entropy will be calculated in states corresponding to conformal vacua of the CFT matter sector, by using the anomalous transformation properties of the CFT.  From the perspective of the dilaton gravity sector (which is not conformally invariant) these ``vacua'' correspond to excitations above the ground state.  These excitations can carry nonzero energy and entropy across horizons, providing a nontrivial check of the GSL.

The action of a generalized dilaton theory \cite{dilrev} coupled to matter is given by
\begin{equation}\label{GDT}
\mathcal{S} = \int d^2x \sqrt{-g} \left[ \frac{R}{2}X + \frac{U(X)}{2}(\nabla X)^2 + V(X) + \mathcal{L}_\mathrm{matter}(X,\,\Phi) \right],
\end{equation}
where $X$ is the dilaton field, and $\Phi$ is a set of matter fields.

A specific example is spherically reduced gravity, in which the dilaton $X$ may be taken to be proportional to the spherical area $A$.  For example, the spherically symmetric sector of $3 + 1$ general relativity (GR) is described by the following action \cite{dilrev}:
\begin{equation}
\mathcal{S} = \int d^2x \sqrt{-g} \left[ \frac{A}{16 \pi G}
\left( R + \frac{1}{2} A^{-2} (\nabla A)^2 \right) - \frac{1}{2G} +
\mathcal{L}_\mathrm{matter}(A,\,\Phi)\right].
\end{equation}
This theory obviously permits black hole solutions, coming from the four dimensional Schwarzschild solution.  By dimensional reduction the generalized entropy of the black hole will be (in units with $\hbar = 1$):
\begin{equation}\label{gen}
S_\mathrm{gen} = \frac{\langle A \rangle}{4G} + S_\mathrm{out}.
\end{equation}
The first term is the Bekenstein-Hawking entropy of the horizon, proportional to the dilaton-area $A$.  In accordance with the arguments of Ref. \cite{10proofs}, we take the expectation value of $A$ so that $S_\mathrm{gen}$ will be a number rather than an operator.  The second term is the entropy of matter fields outside of the horizon, given by a suitable renormalization of the divergent von Neumann entropy $-\mathrm{tr}(\rho\,\ln\,\rho)$.

In a semiclassical analysis, one quantizes the matter sector and couples its expectation value to the gravitational sector of the theory.  In order to justify this approximation in the context of horizon thermodynamics, one needs either a large number of matter fields, or else weak gravitational fields \cite{10proofs}.  The latter option will be selected in this paper, and implemented formally by assuming that $G \ll A$ on the black hole horizon.  Higher order effects of the metric or dilaton on the matter fields can be neglected.

The results of this article will apply to all generalized dilaton theories coupled to any conformal matter sector.  However, without loss of generality, we will consider the special case in which a) the gravitational action is spherically reduced GR, and b) $\mathcal{L}_\mathrm{matter}$ does not depend on the dilaton field $X$.  Assumption (a) is made so as to make use of the normal variables of 4 dimensional GR, in which the dilaton is the area.  However, the restriction is without loss of generality as explained at the end of section \ref{grav}.  Assumption (b) is made for simplicity of exposition; without it, the matter sector might have to be described by a one-parameter family of CFT's depending on the value of the dilaton field $X$.  However, on a stationary horizon, the value of the dilaton field is constant.  Since this article is only concerned with small perturbations to such stationary horizons, the coupling of the matter to the dilaton is irrelevant.

Our strategy is as follows: Since the matter sector is a CFT, it transforms under to an infinite dimensional symmetry group.  Because of the conformal anomaly, the quantum vacuum is not invariant under general conformal transformations.  This leads to an infinite family of states whose stress-energy tensor $T_{ab}$ and outside entropy $S_\mathrm{out}$ are easily calculable.  The GSL can then be tested in these states, based on the (nonconformal) gravitational effect of the stress-energy tensor on the area $A$.

The plan of this article: section \ref{grav} describes the classical gravitational aspects of the theory, section \ref{flat} describes the entropy anomaly in flat spacetime and explains why the entropy is not invariant under local Lorentz boosts if the central charges $c$ and $\tilde{c}$ are unequal.  Section \ref{horizon} derives the anomalous transformation properties of the generalized entropy under conformal reparameterizations of a causal horizon.  It introduces the ``entropic focusing'', a second derivative of the entropy which transforms as a primary quantity under conformal symmetry.  Section \ref{second} uses these transformation properties to derive the GSL for causal horizons.  Section \ref{ADH} demonstrates the failure of the GSL for apparent and dynamical horizons, as well as any other definition of the horizon that differs from the causal horizon in how it responds to a flux of null energy.  Finally, section \ref{prosp} discusses whether the GSL might be a consequence of a more local thermodynamic principle, valid on all null surfaces.

\section{Gravitational Aspects}\label{grav}

In the uncompactified 4-dimensional Einstein theory, the entropy production of a horizon is proportional to the expansion $\theta = A^\prime / A$, where prime means the derivative with respect to an affine parameter $\lambda$.  Furthermore the uncompactified theory should obey the Raychaudhuri equation (written with use of the Einstein equation):
\begin{equation}\label{Ray}
\theta^\prime = -\frac{\theta^2}{2} - (8\pi G)\,^{(4)}T_{ab} k^a k^b,
\end{equation}
where the shear term is absent because of spherical symmetry, and the stress-energy of the CFT is considered to be uniformly distributed along the compactified dimensions.  The difference in normalization between the four dimensional and two dimensional stress-energy tensors is therefore
\begin{equation}\label{Ts}
2\pi \,^{(4)}T_{ab} k^a k^b =\,^{(2)}T_{ab} k^a k^b / A,
\end{equation}
where the factor of $2\pi$ comes from the fact that the uncompactified $^{(4)}T$ is normalized according to GR conventions, while $^{(2)}T$ is normalized according to string conventions.  Using Eqs. (\ref{Ray}) and (\ref{Ts}), one finds that for small matter perturbations (for which nonlinear terms in $\theta$ can be neglected), the two dimensional theory satisfies
\begin{equation}\label{linray}
A^{\prime\prime} = -4G\,^{(2)}T_{ab} k^a k^b.
\end{equation}
Since the gravitational part of the Lagrangian is not conformal, this equation is not invariant under nonaffine reparameterizations of $\lambda$.  In its equilibrium (Hartle-Hawking) state, the horizon has $T_{ab} k^a k^b = 0$ everywhere, and $S_\mathrm{out} = \mathrm{const.}$

Eq. (\ref{Ray}) is the Euler-Lagrange equation associated with varying the $g_{ab} k^a k^b$ component of the metric on the horizon.  If we had instead started with the generalized dilaton theory given by Eq. (\ref{GDT}), the first term gives a contribution proportional to $\ddot{X}$, the second term gives a contribution proportional to $U(X)(\dot{X})^2$, and the potential term gives no contribution at all.  Only the first term contributes to the linearized equation.  This means that neither $U$ nor $V$ will contribute to Eq. (\ref{linray}).  Therefore all generalized dilaton theories have the same linearized horizon thermodynamics properties as spherically reduced GR.

\section{The Entropy Anomaly in Flat Spacetime}\label{flat}

Suppose that you have two entangled quantum mechanical systems, described by Hilbert spaces $\mathcal{H}_1$ and $\mathcal{H}_2$.  The Hilbert space of the whole system is $\mathcal{H}_1 \otimes \mathcal{H}_2$, and let the system be in a pure state $\Psi$.  If you wanted to calculate the density matrix state $\rho$ of system 1, you would trace over the degrees of freedom in $\mathcal{H}_2$:
\begin{equation}\label{restrict}
\rho = \mathrm{tr}_2 |\Psi \rangle \langle \Psi |.
\end{equation}
You could then measure the amount of entanglement using the von Neumann entropy:
\begin{equation}
S = -\mathrm{tr}(\rho\,\ln\,\rho).
\end{equation}
In quantum field theory, things are more complicated, because the entanglement entropy of any bounded region of space is divergent.\footnote{Technically, in each region there still exists an algebra of observables in the region, and there still exist mixed states $\rho$, defined as linear functionals on the space of observables.  But there no longer exists a Hilbert space; hence $\rho$ can no longer be viewed as a density matrix.} In the particular case of a 1+1 CFT, the entanglement entropy in any interval $(x,\,y)$ is logarithmically divergent at both endpoints.

In order to render the vacuum entropy finite, one must impose an ultraviolet cutoff which eliminates contributions to the entropy from high frequency field modes.  There are a variety of cutoffs that one might use.  One possible choice \cite{discrete} is to approximate the quantum field theory by a lattice theory in which there is a minimum spatial length $\epsilon$.  In this case, there will be some specific vacuum state $\Psi$ which minimizes the Hamiltonian, and the density matrix $\rho$ in an interval can be obtained using Eq. (\ref{restrict}).

A second possible method \cite{lnZ} is to find a Hamiltonian flow with respect to which the state $\rho$ of the interval is a thermal state.  In a CFT, this can always be done by using conformal symmetry to map the interval $(x,\,y)$ to the interval $(0,\,+\infty)$ which is thermal because of the Unruh effect.  Formally, this means that
\begin{equation}
\rho = \frac{e^{-\beta H}}{Z},
\end{equation}
where $H$ is the Hamiltonian and $Z = \mathrm{tr}(e^{-\beta H})$ is the partition function.  Although $Z$ diverges, formally it is an identity that
\begin{equation}
S = -\mathrm{tr}(\rho\,\ln\,\rho) = \left( 1 - \frac{d}{d\beta} \right) \ln Z(\beta).
\end{equation}
One can then interpret $Z$ as the partition function of a path integral over a Euclidean manifold with conical singularities at the endpoints of the interval.  This allows one to use standard quantum field regulators for $Z$ as a way to define the entanglement entropy $S$.

A third way \cite{mutual} to regulate the entropy $S$ is to use the mutual information.  For any two systems $1$ and $2$, the mutual information is defined as
\begin{equation}
I_{12} = S_1 + S_2 - S_{12}.
\end{equation}
In the special case where the joint system is pure, $S_{12} = 0$ and $S_1 = S_2$, and thus the entanglement entropy is
\begin{equation}
S = I/2.
\end{equation}
Therefore the entanglement entropy is formally equivalent to the mutual information.  However, unlike the entropy, the mutual information is finite for any pair of regions separated by a nonzero minimum proper distance $\epsilon$ (in some frame of reference).  Hence, by letting system $1$ be the fields inside the interval $(x,\,y)$, and system $2$ be the fields outside the interval $(x - \epsilon, \, y + \epsilon)$, one obtains a finite regularization of the entanglement entropy.  This procedure has the advantage that it can be used to calculate the entropy of other states besides the vacuum state.

In conformal theories, the coefficients of logarithmic divergences are \emph{universal}, i.e. they have to be the same regardless of which regulator is chosen.  To see this, consider two different procedures for regulating the entropy, $S(\epsilon)$ and $S^\prime(\epsilon)$, $\epsilon$ being the characteristic distance scale of each cutoff.  Let each entropy be logarithmically divergent for small values of $\epsilon$:
\begin{eqnarray}
S(\epsilon) = B - C \ln \epsilon, \\
S^\prime(\epsilon) = B^\prime - C^\prime \ln \epsilon
\end{eqnarray}
By virtue of scaling invariance:
\begin{equation}
S(\epsilon) - S^\prime(\epsilon) = S(a\epsilon) - S^\prime(a\epsilon) = [S(\epsilon) - C \ln a] - [S^\prime(\epsilon) - C^\prime \ln a],
\end{equation}
which implies that $C = C^\prime$.

\begin{figure}[ht]
\centering
\includegraphics[width=.6\textwidth]{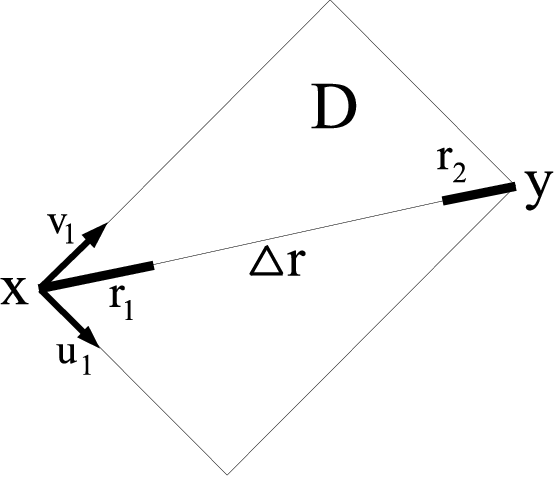}
\caption{\footnotesize A spacetime interval $(x,\,y)$ of proper length $r$, lying inside of a parallelogram representing its causal domain $D$.  There is an infinite entanglement entropy in $D$, or equivalently on $(x,\,y)$.  $r_1$ and $r_2$ are the length-scales associated with UV cutoffs.  If one wishes to consider the effects of boosting the cutoff length, one can also view the cutoff as a vector with null components $u_1$ and $v_1$.}\label{interval}
\end{figure}

In any CFT, this universal entropy divergence turns out to be related to the left and right moving central charges $c$ and $\tilde{c}$ \cite{discrete, lnZ, mutual}:\footnote{If $\mathcal{L}_\mathrm{matter}$ were to depend on the dilaton $A$, then the central charges $c(A)$ and $\tilde{c}(A)$ might depend on location.}
\begin{equation}\label{log}
S = -\mathrm{tr}(\rho\,\ln\,\rho) =
\frac{c + \tilde{c}}{12} \ln \left( \frac{\Delta r^2}{r_1 r_2} \right) + s(\rho).
\end{equation}
Here $\Delta r$ is the proper length of the interval $(x,\,y)$, $r_{1,2} \ll \Delta r$ are ultraviolet distance cutoffs on the left and right sides of the interval (Figure \ref{interval}), imposed in the frame of reference of the interval $(x,\,y)$.  $s$ is the UV convergent contribution to the entropy, which depends on the state $\rho$ and the regulator.

Although $s$ has no ultraviolet divergences, it may still have infrared divergences.  For example, in the case of the massless free scalar field $\Phi$, there is a constant infrared divergence in the entanglement entropy in any interval, due to quantum fluctuations whose wavelength is long compared to the interval \cite{FPST}.  This is because of the translation symmetry $\Phi \to \Phi + a$ of the scalar field.  On the other hand, there is no infrared divergence associated with a massless free fermion on an interval.  This article will be primarily concerned with the case in which one of the endpoints of the interval is taken to infinity.  In this case there is an additional IR logarithmic entropy divergence at long distances.  Assuming that this infrared divergence is cut off at the same very long distance at all times, this divergence just shifts $S$ by a constant factor and thus makes no difference to the GSL (which is only concerned with changes in $S$).

The CFT is conveniently analyzed in terms of null coordinates $u$ and $v$, in which the proper distance in flat spacetime can be written as
\begin{equation}
(\Delta r)^2 = -\Delta u \Delta v.
\end{equation}
The central charge $c$ has to do with left-moving fields, which depend only on the coordinate $v$.  Similarly, $\tilde{c}$ has to do with right-moving fields, which depend only on $u$.  Using this additional piece of information, it is possible to write the entropy divergence in a way which distinguishes the roles of $c$ and $\tilde{c}$:
\begin{equation}\label{uv}
S = \frac{c}{12} \ln \left( \frac{(\Delta v)^2}{v_1 v_2} \right)
+ \frac{\tilde{c}}{12} \ln \left( \frac{(\Delta u)^2}{u_1 u_2} \right)
+ s(\rho),
\end{equation}
where $\Delta v$ is the difference between the $v$ coordinate values of the endpoints, $v_1$ and $v_2$ are the $v$-coordinate lengths of the UV cutoff, and the same for $u$.

To derive Eq. (\ref{uv}), note that the entropy divergence must split into two parts: a left moving entropy $S_L$ associated with the left-movers, and a right moving part $S_R$ associated with the right-movers.  This means that for any given interval, the entropy can be written as
\begin{equation}
S = S_L(c, v_1, v_2, \Delta v) + S_R(\tilde{c}, u_1, u_2, \Delta u) + s(\rho).
\end{equation}
In the special case where the UV cutoff vectors lie in the same line as the interval, all $u$ and $v$ intervals are in the same ratio as $r$ intervals:
\begin{equation}\label{ratios}
u_1 : u_2 : \Delta u :: v_1 : v_2 : \Delta v :: r_1 : r_2 : \Delta r,
\end{equation}
and Eq. (\ref{log}) may be used to determine that
\begin{equation}\label{SL}
S_L = \frac{c}{12} \ln \left( \frac{(\Delta r)^2}{r_1 r_2} \right)
= \frac{c}{12} \ln \left( \frac{(\Delta v)^2}{v_1 v_2} \right),
\end{equation}
while
\begin{equation}\label{SR}
S_R = \frac{\tilde{c}}{12} \ln \left( \frac{(\Delta r)^2}{r_1 r_2} \right)
= \frac{\tilde{c}}{12} \ln \left( \frac{(\Delta u)^2}{u_1 u_2} \right).
\end{equation}
However, since $S_L$ is not a function of $u_{1,2}$ or $\Delta u$, Eq. (\ref{SL}) must actually be true even when Eq. (\ref{ratios}) is not true.  Similarly, since $S_R$ is not a function of $v_{1,2}$ or $\Delta v$, Eq. (\ref{SR}) is also independent of Eq. (\ref{ratios}).  This shows that Eq. (\ref{uv}) holds even when the cutoffs are boosted relative to the interval $(x,\,y)$.

\begin{figure}[ht]
\centering
\includegraphics[width=.9\textwidth]{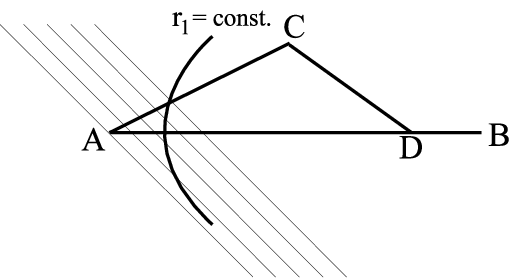}
\caption{\footnotesize The spatial slices $ACDB$ and $ADB$ contain exactly the same information as each other.  So naively one would expect that they also contain the same (renormalized) entanglement entropy $S$.  But consider e.g. a left-moving chiral field, for which $c > \tilde{c} = 0$.  All information travels to the left at the speed of light.  After cutting off the entanglement entropy at a fixed proper distance $r_1$ from point $A$, one finds $S(ACDB) < S(ADB)$ because some of the entropy has propagated leftward past the cutoff.  Since the entropy depends on the boost angle at which it is measured, it fails to be invariant under local Lorentz symmetry.  The opposite sign entropy change would occur for right-moving fields.  When $c = \tilde{c}$, the entropy is the same in all reference frames.}\label{anomaly}
\end{figure}

The theory with $c \ne \tilde{c}$ is sick once gravitational effects are taken into account because of the trace anomaly \cite{trace}.  Interestingly, even in flat spacetime, there is a problem for thermodynamics, because the total entropy of an interval in Eq. (\ref{uv}) changes when the cutoffs e.g. $u_1$ and $v_1$ are boosted (Fig. \ref{anomaly}).  This means that there is no way to define the generalized entropy of an interval using a Lorentz invariant, local cutoff.

Thus we shall require $c = \tilde{c}$ in what follows (this is always true if the CFT comes from dimensional reduction).  This means that the entropy only depends on the proper lengths $r_{1,2}$ of the cutoffs, not on the boost angle of the cutoffs.

There is still, however, a scale anomaly which makes the regulated entropy change under a conformal rescaling of either endpoint.  Since any cutoff must neglect all of the entropy which is localized sufficiently close to the cutoff and count all the entropy which is sufficiently far away, one can understand this anomaly as coming from the fact that any rescaling of the endpoint will move entropy from one region to the other.  This however, is not a conceptual problem since the theory taken as a whole (including the gravitational sector) depends on the conformal factor in the metric.

In order to be consistent with the symmetries of the full gravitational theory, we therefore choose the cutoffs to be at a fixed (but small) proper distance with respect to the metric $g_{ab}$.  This choice of cutoff means that any time one performs a Weyl rescaling of the metric, one also must adjust the coordinate size of the cutoffs, and hence change the entropy as well.  Thus the entropy depends functionally not only on the state and the choice of endpoints, but also on the conformal factor of the metric at each endpoint:
\begin{equation}
S(x,\,y,\,g(x),\,g(y),\,\rho),
\end{equation}
although for convenience the last three arguments will be left implicit below.

Using Eq. (\ref{uv}), it is then possible to write down the effect of a conformal transformation on the entropy of an interval.  The interval is defined by its two endpoints $(x,\,y)$.  An ``active'' conformal transformation acts in the following way:
\begin{enumerate}
\item Each point $x$ on the spacetime is moved to a new point $f(x)$, where $f$ is an active diffeomorphism that preserves the causal structure of the Minkowski metric $\eta_{ab}$;
\item The resulting metric is multiplied by the Weyl rescaling of the metric $\Omega(x)$, $g_{ab} \to \Omega^2(x) g_{ab} = \eta_{ab}$ ($x$ being the old coordinate location of the points), restoring the original metric and the original cutoffs.
\item The quantum fields in a state $\rho$ are dragged by the diffeomorphism to new locations, transforming the state to a new state $\sigma$;
\item The interval $(x,\,y)$ remains in place at its old coordinate location.
\end{enumerate}
Because the fields move relative to the interval and the cutoff, the entropy $S_f(x,\,y)$ of the conformally transformed fields need not equal the entropy $S(x,\,y)$ of the untransformed fields.  We will calculate its transformation law with a two-step procedure.

First of all, if one acts on the fields \emph{and the cutoffs} with the diffeomorphism $x \to f(x)$, the entropy can be calculated simply by displacing the endpoints of the interval:
\begin{equation}\label{step1}
S_{\mathrm{Diff}}(f(x),\,f(y)) = S(x,\,y),
\end{equation}
where the new cutoffs $u_{1,2}^\prime$, $v_{1,2}^\prime$ are related to the old ones by
\begin{equation}
u_1^\prime v_1^\prime = \Omega^2(x)\,u_1 v_1, \qquad u_2^\prime v_2^\prime = \Omega^2(y)\,u_2 v_2.
\end{equation}
This differs from the conformal transformation in step 5, since the conformal transformation does not modify the cutoffs.

In the second step, one restores the cutoffs to their original values using Eq. (\ref{uv}), in order to find the conformal transformation of the entropy:
\begin{equation}\label{Sln}
S_f(f(x),\,f(y)) = S(x,\,y) + \frac{c}{12} \ln(\Omega^2(x)\,\Omega^2(y)).
\end{equation}
This transformation law can also be expressed in infinitesimal form by defining $f = x + \delta x$, where $\delta x \equiv \xi$ is a conformal Killing vector:
\begin{equation}\label{inf}
\delta S(x,\,y) =
- \left[ \xi^a(x) \frac{\partial}{\partial x^a}
+ \xi^a(y)\frac{\partial}{\partial y^a} \right] S(x,\,y)
+ \frac{c}{12} \left[ \frac{\partial \xi^a(x)}{\partial x^a} + \frac{\partial \xi^a(y)}{\partial y^a} \right],
\end{equation}
using the fact that
\begin{equation}
\delta \Omega^2 = \frac{\partial \xi^a(x)}{\partial x^a}.
\end{equation}

\section{Conformal Transformations on the Horizon}\label{horizon}

Although the discussion thus far has been in flat spacetime, the same local divergence structure must also appear at each point in curved spacetime, because the curvature does not matter locally.  Hence Eq. (\ref{inf}) also applies in curved spacetime.\footnote{The stress-energy tensor of the CFT also has a trace anomaly associated with the curvature at each point: $T^a_a = T_{uv} g^{uv} = (c/12)R$.  The trace anomaly breaks the local conformal symmetry of the spacetime, and can cause energy to be transferred from left-moving to right-moving fields.  Thus $T_{vv}$ becomes a function of $u$ as well as $v$.  Since the following calculation takes place entirely at $u = 0$, the trace anomaly is not important.}

The anomalous transformation laws will now be applied to states on a dilaton black hole background.  Let the black hole background be stationary, up to a small semiclassical perturbation of the CFT fields.  As a result of this perturbation the black hole will no longer contain a Killing horizon.  Nevertheless it still contains a \emph{future event horizon}, defined as the boundary of the past of asymptotic future null infinity $\mathcal{I}^+$.  In other words, the exterior of the event horizon is the region from which one can escape from the black hole.  This future event horizon is also an example of a future causal horizon, which is defined as the boundary of the past of any future-infinite timelike worldline.\footnote{Rindler and de Sitter horizons are examples of causal horizons which are not event horizons.  The GSL seems to apply to general causal horizons \cite{JP03}.  The methods used in this article could be applied to these kinds of causal horizons as well, but for concreteness the dilaton black hole has been selected.}  Since the horizon is defined using the causal structure of the spacetime, it is always a null surface (in 1+1 dimensions, a null curve).  As a causal horizon, it has two important properties: 1) it is always a null surface, and 2) because it is a ``future'' causal horizon, it satisfies the future boundary condition $\theta(\lambda = +\infty) = 0$. 

Without loss of generality, we choose null coordinates such that $u = 0$ is the causal horizon, while $v$ is an affine parameter on the horizon.  In the stationary vacuum (Hartle-Hawking) state, the dilaton field $A$ and the entropy $S_\mathrm{out}$ must be constant with respect to $v$; thus the GSL holds in this state. 

$S_\mathrm{out}$ receives contributions from left-moving fields (which fall across the horizon) and right-moving fields (which escape to $\mathcal{I^+}$).  The left and right moving fields are unentangled in the Hartle-Hawking state.  The constancy of the generalized entropy is due to a balance between left-moving modes disappearing across the horizon, and right-moving modes being redshifted out of the entropy cutoff.

We seek to test the GSL by applying an active conformal transformation.  Since a transformation of the $u$ coordinate is translation invariant in the $v$ direction, only left-moving conformal transformations of $v$ are interesting.  The transformation properties of the entropy and energy determine the generalized entropy of the transformed state.  \emph{Each of these conformally transformed states will be shown to obey the GSL.}

Notational simplifications: Below, all tensors will be evaluated at $u = 0$; indices will be suppressed because they are all $v$ components.  $X^\prime$ will mean the derivative of $X$ in the $v$-direction.  Furthermore the outside entropy of the CFT fields $S_\mathrm{out} = S(v,\,+\infty)$ will be written simply as $S$ (we need not concern ourselves with the IR entropy regulator since it will not be a function of $v$).  Also, the $f$ subscript for the transformed entropy will be dropped since it can be presumed for all quantities $X(f)$ which are written as functions of $f$.  Finally, expectation value signs around quantities such as $A$ and $T$ will be presumed. 

From Eq. (\ref{inf}), the transformation law of $S$ is
\begin{equation}\label{S}
\delta S = -\xi S^\prime + \frac{c}{12} \xi^\prime.
\end{equation}
The first term in Eq. (\ref{inf}) (which takes the same general form for all quantities) is the result of translating the fields relative to the endpoint, while the second term comes from dilating the fields relative to the cutoff.  Take the derivative of Eq. (\ref{S}):
\begin{equation}\label{Sp}
\delta S^\prime = -\xi S^{\prime\prime} - \xi^\prime S^\prime
+ \frac{c}{12}\xi^{\prime\prime}.
\end{equation}
Eq. (\ref{Sp}) can be integrated to find the transformation properties under a finite active conformal transformation sending $v \to f(v)$:
\begin{equation}\label{SpI}
S^\prime(f) = (f^\prime)^{-1} \left[
S^\prime(v) + \frac{c}{12}\frac{f^{\prime\prime}}{f^\prime}
\right].
\end{equation}
A second derivative of the entropy yields
\begin{equation}\label{Spp}
\delta S^{\prime\prime} = -\xi S^{\prime\prime\prime}
- 2\xi^\prime S^{\prime\prime} - \xi^{\prime\prime}S^\prime
+ \frac{c}{12}\xi^{\prime\prime\prime}.
\end{equation}
This transformation law is hard to integrate because of the dependence on $S^\prime$.  Conformal transformations act in a non-affine manner on the two-dimensional vector space $(S^\prime,\,S^{\prime\prime})$.  A somewhat nicer quantity is
\begin{equation}
\$ = S^{\prime\prime} + \frac{6}{c}(S^\prime)^2,
\end{equation}
which characterizes the curves in $(S^\prime,\,S^{\prime\prime})$ which are invariant under the action of $\xi^{\prime\prime}$.  This quantity obeys the transformation law
\begin{equation}
\delta \$ = -\xi \$^\prime - 2\xi^\prime \$ + \frac{c}{12}\xi^{\prime\prime\prime}.
\end{equation}
This transformation law looks suspiciously similar to the transformation law of the stress-energy tensor $T_{vv}$ \cite{ginsparg} (normalized in string units):
\begin{equation}\label{Tlaw}
\delta T = -\xi T^\prime - 2\xi^\prime T + \frac{c}{12}\xi^{\prime\prime\prime}
\end{equation}
Therefore their difference
\begin{equation}\label{Ldef}
L = T - \$ = T - S^{\prime\prime} - \frac{6}{c}(S^\prime)^2
\end{equation}
remarkably is primary under conformal transformations:
\begin{equation}
\delta L = -\xi L^\prime - 2\xi^\prime L,
\end{equation}
(i.e. its transformation law does not depend on more than one derivative of $\xi$) and therefore transforms like a weighted tensor under a finite conformal transformation:
\begin{equation}\label{LI}
L(f) = (f^\prime)^{-2} L(v).
\end{equation}
Together with Eq. (\ref{SpI}), Eq. (\ref{LI}) provides an easy way to calculate how $S$ and $T$ change under a general conformal transformation.

\section{The Second Law on the Causal Horizon}\label{second}

The next step is to see how the generalized entropy $S_\mathrm{gen}$ changes under a conformal transformation.  Because of the teleological boundary conditions used to define the horizon, $S_\mathrm{gen}$ does not transform locally---it depends on what is going to happen in the future.  But the second derivative of $S_\mathrm{gen}$ does transform locally:
\begin{equation}
S_\mathrm{gen}^{\prime\prime} = \frac{A^{\prime\prime}}{4G} + S^{\prime\prime}
= - T + S^{\prime\prime} = -(L + \frac{6}{c}(S^\prime)^2),
\end{equation}
where the linearized Raychaudhuri Eq. (\ref{linray}) has been used to relate $A$ to $T$.  By imposing a stationary final boundary condition on the event horizon:
\begin{equation}\label{futb}
S_\mathrm{gen}^{\prime}|_{+\infty} = 0,
\end{equation}
one can express the GSL at a horizon point $X$ as the following condition:
\begin{equation}
S_\mathrm{gen}^\prime(X) =
-\int_X^{+\infty} S_\mathrm{gen}^{\prime\prime}\,dv
= \int_X^{+\infty} (L + \frac{6}{c}(S^\prime)^2)\,dv \ge 0.
\end{equation}
The second term is automatically nonnegative and therefore only helps the GSL to be satisfied, while a sufficient (but not necessary) condition for the first term to be nonnegative is that
\begin{equation}
L \ge 0
\end{equation}
everywhere on the horizon to the future of $X$.  This condition has two advantages: 1) it is localized with respect to $v$ on the horizon, and 2) it is invariant under conformal transformations.  One could regard $L$ as a kind of ``entropic focusing'', whose positivity ensures the validity of the GSL.  It is similar to how the positive focusing of the area given by the Raychaudhuri Eq. (\ref{Ray}) ensures the validity of the classical second law \cite{hawking71}.

The Hartle-Hawking vacuum state is stationary with respect to Killing time translations.  This symmetry (together with continuity across the bifurcation surface) implies that the dilaton $A$ and the entropy $S$ are constant with respect to $v$.  Since $L$ consists of derivatives of $A$ and $S$, it vanishes in the Hartle-Hawking state.  It follows that any general conformal transformation of the vacuum state also has $L = 0$ and hence also obeys the GSL.

Furthermore, if the field theory is free, one can build coherent states which approximate classical states in the theory.  A coherent state may obtained by acting on the vacuum state $\Omega$ with the following transformation of the free fields $\phi$:
\begin{equation}\label{trans}
\phi (x) \to \phi (x) + \phi_c (x),
\end{equation}
where $\phi_c$ is a \emph{classical} solution to the free equations of motion.  Because this transformation acts as a unitary transformation of the state, the von Neumann entropy $-\mathrm{tr}(\rho\,\ln\,\rho)$ is unaffected by this transformation.  Consequently the entanglement entropy of a coherent state is the same as the entanglement entropy of the vacuum:
\begin{equation}\label{Ss}
S(\Psi) = S(\Omega).\footnote{In order to make this argument completely precise, one also has to show that the field transformation ``symmetry'' (\ref{trans}) is compatible with the regulator used to make the von Neumann entropy finite (cf. section \ref{flat}).  Since neither the lattice nor the mutual information regulator break this symmetry in any way, the renormalized entropy also satisfies Eq. (\ref{Ss}).}
\end{equation}
Since the stress-energy $T$ is a normal-ordered quadratic function of the field operators, one can also calculate the expectation value of $T$ by using a binomial expansion in the free fields.  For example, in a free scalar field theory, $T = :\nolinebreak\!(\phi^\prime)^2\nolinebreak\!:$, and the expected stress-energy of a coherent state $\Psi$ is
\begin{equation}
\langle T \rangle_\Psi = \langle :\!(\phi^\prime)^2\!: \rangle_\Omega
+ 2 \langle \phi^\prime \rangle_\Omega \phi_c^\prime
+ (\phi_c^\prime)^2.
\end{equation}
However, the first term vanishes because the vacuum state has zero energy, while the second (interference) term vanishes because the free field vacuum has $\phi \to -\phi$ symmetry.  Thus in any coherent state, the expected stress-energy tensor is given by the stress-energy of the classical state which it approximates:
\begin{equation}
\langle T \rangle_\Psi =  T(\phi_c),
\end{equation}
which is typically positive for a stable theory.  Thus coherent states have $T \ge 0$ and $S^\prime = 0$, and hence $L \ge 0$.  This means that general conformal transformations of coherent states will also obey the GSL.


These states are far from being the only states in the theory, but they are a nontrivial infinite-dimensional space of states.  The states may be rapidly changing with time.  Recently, the GSL was proven for rapidly evolving semiclassical perturbations to stationary horizons \cite{null}.\footnote{Ref. \cite{null} assumed the existence of a renormalization scheme for the generalized entropy with certain properties.  It also assumed the existence of an algebra of observables measurable on the horizon.  These assumptions should apply to 1+1 CFT's.}  This calculation confirms the proof for these states.

\section{Other Kinds of Horizons}\label{ADH}

Above, we have tested the GSL for the lightlike horizon located at $u = 0$.  However, it is also possible to test the GSL for other kinds of future horizons mentioned in the literature.  It will turn out that the GSL does \emph{not} hold for apparent, trapping, dynamical, or any other kind of horizon besides the causal horizon.  This is because only the causal horizon satisfies the (linearized) Raychaudhuri Eq. (\ref{linray}).

We now review some definitions of different kinds of horizons.  It was observed in section \ref{horizon} that the $u = 0$ curve of the dilaton black hole meets both of the following definitions:
\begin{description}
\item[Event Horizon:] The boundary of the past of asymptotic null infinity $\mathcal{I}^+$.
\item[Causal Horizon:] The boundary of any future-infinite timelike worldline \cite{JP03}.
\end{description}
However, these definitions have been criticized \cite{FPST, hayward, ashtekar} because of their nonlocal ``teleological'' character: the location of a causal horizon depends on how much matter is going to fall across the horizon in the future.

An alternative family of definitions tries to define the location of the horizon based on more local features of the spacetime.  These definitions depend on Penrose's notion of a \emph{trapped surface}, which is a closed $D - 2$ dimensional spacelike surface such that their future-outwards normal lightrays have negative expansion $\theta = (1/A)(dA / d\lambda) < 0$ everywhere.  A \emph{marginally} trapped surface instead has $\theta = 0$ everywhere.  (In the present case of $D = 2$ dilaton gravity, although $\Sigma$ is just a point, one can still define trapped surfaces using the dilaton $A$ as the ``area''.)  One can now define horizons as follows:
\begin{description}
\item[Apparent Horizon] has two distinct definitions in the literature \cite{booth}: 1) Given a foliation of spacetime into asymptotically flat $D - 2$ dimensional spacelike surfaces $\Sigma_t$, for any time `t' the apparent horizon is the outermost boundary of all trapped surfaces on $\Sigma_t$ \cite{hawkingellis}.  Assuming this surface is smooth, this is equivalent to 2) the outermost marginally trapped surface.  Both definitions are also equivalent in dilaton gravity.
\item[(Outer) Trapping Horizon:] A $D - 1$ dimensional surface foliated by margin\-ally trapped surfaces, such that the expansion $\theta$ is positive for null surfaces just outside the horizon, and negative for null surfaces just inside \cite{hayward}.  (Unlike the apparent horizon, this definition refers only to the geometry near the horizon, and therefore does not require asymptotic flatness.)
\item[Dynamical Horizon:] A $D - 1$ dimensional surface foliated by marginally trapped surfaces, for which the \emph{infalling} lightrays are contracting ($\theta < 0$) \cite{ashtekar}.
\end{description}
The trapping horizon \cite{hayward} and the dynamical horizon \cite{ashtekar} can each be shown to obey local forms of the classical first and second laws of black hole mechanics.  Below it will be shown that they do not obey the quantum second law (the GSL).

Specializing to the case of 1+1 dilaton black holes (the dimensional reduction of Schwarzschild), the trapping horizon is a curve all of whose points satisfy
\begin{equation}\label{app}
\theta_v = \frac{1}{A}\frac{dA}{dv} = 0,
\end{equation}
and also the inequality
\begin{equation}\label{outer}
\partial_u \theta_v < 0
\end{equation}
which ensures that the horizon is an outer one rather than an inner one.  Since this is the only future outer trapping horizon, it is also the apparent horizon.

In the case of the ``dynamical horizon'', the inequality is replaced with
\begin{equation}\label{dyn}
\theta_u = \frac{1}{A}\frac{dA}{du} < 0,
\end{equation}
which ensures that the ingoing lightrays are converging.  Since Eq. (\ref{outer}) and Eq. (\ref{dyn}) are strict inequalities, if they are satisfied at $u = 0$ then they must be satisfied for small positive or negative values of $u$ as well.

Since the 1+1 dimensional dilaton black hole satisfies both inequalities above the bifurcation surface, the apparent, trapping, and dynamical horizons all coincide there.  For brevity we will call it the apparent horizon.

The apparent horizon is stable under small perturbations.  If a small amount of stress-energy falls across the causal horizon at $u = 0$, there will be a nonzero value of $\theta$ on the $u = 0$ surface.  By virtue of the outer condition (\ref{outer}), the apparent horizon can continue to satisfy Eq. (\ref{app}) by moving slightly inwards or outwards to a nonzero $u(v)$ value.  When there is a positive (negative) null stress-energy falling across the apparent horizon, it becomes a spacelike (timelike) surface.  In the spacelike case, there exist classical area-increase theorems \cite{hayward, ashtekar}.

However, once quantum mechanical effects are taken into account, the GSL no longer holds.  Once again, let the Hartle-Hawking equilibrium state be modified by a conformal transformation of the left-moving fields.  Since the left-movers are translation invariant in the $u$ direction, the fact that the apparent horizon moves away from the $u = 0$ surface makes no difference to the entropy $S$ of the matter fields.  However, the area $A$ is now related differently to the infalling stress-energy tensor.  Unlike causal horizons (which satisfy the linearized Raychaudhuri equation $A^{\prime\prime} = -4GT$), the apparent horizon satisfies the first law
\begin{equation}
A^{\prime} = \frac{4G}{\kappa}T
\end{equation}
even when $T$ is rapidly changing.  Here $\kappa$ is the surface gravity of the black hole:
\begin{equation}
\kappa = \frac{\partial_u \theta_v}{\theta_u}.
\end{equation}
Because $v$ is an affine coordinate rather than the Killing time, the surface gravity $\kappa$ will not be constant; instead it must be proportional to $v^{-1}$.  One can now write the GSL as follows:
\begin{equation}
S_\mathrm{gen}^\prime = \frac{T}{\kappa} + S^\prime \ge 0.
\end{equation}
Hence the GSL transforms locally under conformal transformations, according to Eq. (\ref{Sp}) and Eq. (\ref{Tlaw})  Applying a first order conformal transformation to the stationary vacuum, one obtains
\begin{equation}
\delta S_\mathrm{gen}^\prime =
\frac{c}{12} \left( \xi^{\prime\prime} + \frac{\xi^{\prime\prime\prime}}{\kappa} \right).
\end{equation}
From this it follows that one can obtain a state that violates the GSL on the apparent horizon simply by choosing the second or the third derivative of $\xi$ to be negative.  Depending on the sign of $T$, the GSL can be violated on timelike, spacelike, or lightlike apparent horizons.

This same argument can also be used to show that the GSL is violated on apparent, trapped, or dynamical horizons in higher spacetime dimensions ($D > 2$).  For example, consider a 3+1 Schwarzschild black hole minimally coupled to a scalar field.  The s-wave sector of the scalar field reduces to a 1+1 CFT near the horizon.

This decrease of generalized entropy has already been observed in Appendix B of Fiola et al. \cite{FPST}.  That work rejected causal horizons based on an incorrect argument claiming that the GSL would be violated on causal horizons when one sends in very sharp pulses.  However, this argument did not take into account the thermal atmosphere of the black hole itself  \cite{10proofs}.  Having therefore restricted their attention to the apparent horizon, they concluded that the GSL simply does not hold on small timescales.  The causal horizon, however, obeys the GSL even on short timescales.\footnote{One might suppose that the GSL \emph{ought} to be violated on short time scales due to fluctuations in the entropy.  That would be true if the entropy were defined as $\ln\,N$ where $N$ is the number of microstates per macrostate.  However, this article uses the von Neumann definition of the entropy, which is not subject to such fluctuations \cite{10proofs}.}

In fact, the causal horizon is the \emph{only} definition of the horizon which satisfies the GSL, at the level of linear black hole perturbations.  Suppose someone were to come up with some \emph{other} way of defining the horizon---call this an ``O-horizon''.  The linearized dependence of the area of the O-horizon on $T$ can be written as
\begin{equation}
\delta A(v_1) = \int_{-\infty}^{+\infty} r(v_1, v_2)\,\delta T(v_2)\,dv,
\end{equation}
where $r$ is some linear response function.  (So long as $\theta_u \ne 0$, the area can then be used to find the horizon's position).  Because there exist stationary black hole solutions with arbitrary constant values of $A$, the linear response function is ambiguous under constant shifts in the first argument:
\begin{equation}
r(v_1, v_2) \to r(v_1, v_2) + \Delta A (v_2).
\end{equation}
This ambiguity can be fixed by adopting the convention that $r(v_1, +\infty) = 0$.  In the case of the causal horizon,
\begin{equation}\label{causalr}
r(v_1, v_2) = -4G(v_2 - v_1)\theta(v_2 - v_1),
\end{equation}
where $\theta$ is the Heaviside function.


If the GSL does always hold on the O-horizon, then in general $S_\mathrm{gen}^\prime \ge 0$.   Since the Hartle-Hawking state has $S_\mathrm{gen}^\prime = 0$, the Hartle-Hawking state would then minimize the entropy increase, which implies that any first order variation away from Hartle-Hawking would satisfy $\delta S_\mathrm{gen}^\prime = 0$.  Applying this variational principle to a small conformal transformation, the transformed state must satisfy
\begin{equation}\label{balance}
\frac{\delta A}{4G} = - \delta S,
\end{equation}
assuming that $\xi$ satisfies future boundary conditions so that $\delta S(+\infty) = 0$.  But this area/entropy balance is only possible if $A$ responds in just the right way to a conformal transformation $\xi$.  Consider the following conformal transformation:
\begin{equation}
\xi = -\theta(v_2 - v_1)(v_2 - v_1)^2/2.
\end{equation}
This implies that
\begin{equation}
\delta T(v_1) = (c/12)\delta(v_2 - v_1)
\end{equation}
using Eq. (\ref{Tlaw}), and
\begin{equation}\label{Sresp}
\delta S = -(c/12)\theta(v_2 - v_1)(v_2 - v_1)
\end{equation}
using Eq. (\ref{S}).  But now Eq. Eq. (\ref{balance}) can be used with (\ref{Sresp}) to calculate the response of the area $A$ to a delta function in the stress-energy $T$.  The only way for the GSL to hold is if the response function $r(v_1, v_2)$ is given by Eq. (\ref{causalr}).  So the teleological character of the causal horizon is necessary for the GSL to hold!

\section{Prospects}\label{prosp}

It has been shown that for 1+1 conformal vacua, the GSL holds on the causal horizon of a dilaton black hole, but not on apparent, trapping, or dynamical horizons.  This indicates that the causal definition is the correct choice for black hole thermodynamics.  The result is a special case of the general proof of the GSL for rapidly changing weak semiclassical perturbations, described in Ref. \cite{null}.  (The proof assumes that it makes sense to restrict fields to the horizon, and then uses the null translation and boost symmetries of the horizon to derive the GSL.)

Because of the simplicity of dilaton gravity, it might not be too difficult to generalize this work beyond the weak gravity limit, and thus test the GSL in a new regime.  Such a project would need to take into account the effects of the nonlinear $A^{-1}(\nabla A)^2$ term in the gravitational Lagrangian on the horizon entropy.  Because the expectation value of a product is in general different from the product of the expectation values, in general one would need to go beyond the semiclassical approximation by taking into account dilaton fluctuations (unless there are a large number $N$ of matter fields).

The GSL is the quantum gravitational analogue of the second law of thermodynamics.  Therefore, one expects it to hold for some basic statistical mechanical reasons, coming purely from information theory.  This makes it rather surprising that the definition of the GSL is so nonlocal.  The nonlocality comes in for two different reasons: 1) the causal horizon is teleological, so its existence and location depends on what is going to happen in the future, and 2) the outside entropy term $S_\mathrm{out}$ refers to matter fields arbitrarily far away from the horizon itself.  If either of these conditions is relaxed, the GSL can be violated.\footnote{The necessity of condition (1) was shown in section \ref{ADH} of this article, and is further discussed in Refs. \cite{10proofs} and \cite{cosm}.  The necessity of condition (2) was observed in Ref. \cite{derivative}, and is a consequence of Corollary 1.3 of Ref. \cite{cosm}.}  If the GSL is formulated in such a nonlocal way, can it be a fundamental law of nature?

Although the GSL is nonlocal, it might nonetheless be a logical consequence of some other set of local statistical principles.  For example, in classical general relativity, Hawking's area increase theorem \cite{hawking71} is a nonlocal result, which is nevertheless derived from the local Raychaudhuri focusing equation, through use of the concept of ``trapped surfaces''.  This example is particularly relevant given that the ``Area Increase Theorem'' is just the classical limit of the GSL.  Furthermore, the GSL can be used (at least semiclassically) to prove that there is a quantum analogue of ``trapped surfaces'' (Theorem 4 of Ref. \cite{cosm}.)  So it seems reasonable to cast around for a local, quantum focusing condition which might imply the GSL.

That challenge has been partly met here, using the condition $L \ge 0$, where $L$ is the entropic focusing defined in Eq. (\ref{Ldef})) using two derivatives of the horizon entropy.  It is a sort of entropic analogue of the area focusing equation.  This condition is local in sense (1), i.e. it does not depend on the teleological boundary condition, but holds on all null surfaces.  When one applies it to causal horizons, one also has a future boundary condition given by Eq. (\ref{futb}), and then the GSL applies.

However, all of this has only been shown for conformal vacua (and coherent states).  Also, the quantity $L$ is still nonlocal in sense (2), since it continues to refer to $S_\mathrm{out}$, the entire entropy outside the horizon.

It would be interesting to find out whether $L \ge 0$ holds for more general states.  Perhaps there is even an analogue of $L$ for nonconformal matter sectors, or for theories in higher dimensions.

\small

\subsection*{Acknowledgements}
I am grateful for editing suggestions from Ted Jacobson, William Donnelly, and an anonymous referee.  Supported by NSF grants PHY-0601800, PHY-0903572, the Maryland Center for Fundamental Physics, and the Center for Fundamental Theory at Penn State.

\end{document}